\documentclass[12pt]{article}
\begin{document}

\title{Quantum Weibel instability}
\author{F. Haas \footnote{Electronic mail: ferhaas@tp4.ruhr-uni-bochum.de}}
\date{\relax}
\maketitle
\begin{center}
{Institut f\"ur Theoretische Physik IV, 
Ruhr-Universit\"at Bochum\\ D-44780 Bochum Germany \\
\vskip .5cm
Universidade do Vale do Rio dos Sinos - UNISINOS 
\\ Av. Unisinos, 950, 93022--000 S\~ao Leopoldo RS Brazil}
\end{center}

\begin{abstract}
The Weibel instability is analyzed for quantum plasmas described by the Wigner-Maxwell model. For a suitable class of electromagnetic potentials, 
the Wigner-Maxwell system is linearized yielding a general dispersion relation for transverse electromagnetic waves. For a double Gaussian equilibrium 
with temperature anisotropy, the derived dispersion relation generalizes the classical Weibel instability equation. More detailed analytical results
are obtained for the cases of extreme temperature anisotropy and for a three-dimensional water bag distribution. In all cases, quantum effects tends 
to weaken or suppress the instability. Applications are discussed for dense astrophysical objects like white dwarfs and neutron stars as well as for tunnel-ionized plasmas with controllable perpendicular plasma temperature. 
\end{abstract}

\section{Introduction}
Quantum plasmas have attracted renewed attention in recent years due to the ongoing miniaturization of ultra small electronic devices and 
micro mechanical systems \cite{Markowich}, to the relevance of quantum effects for dense laser-plasmas and micro plasmas \cite{Becker} and for dense 
astrophysical objects \cite{Opher}. Quantum phenomena are relevant for these systems for a variety of reasons, the most usual being the de Broglie 
wavelength of the charge carriers (electrons, positrons, holes and so on) becoming comparable to the characteristic dimensions of the system. Quantum 
ion-acoustic waves \cite{H}, a quantum magnetohydrodynamics model \cite{HH}, shear Alfv\'en modes in ultra-cold quantum magneto plasmas \cite{Shukla2}, quantum corrections for the Zakharov system \cite{Garcia}-\cite{HAA}   
and nonlinear solutions for quantum magneto plasmas \cite{HaasEPL} have been constructed. New quantum modes have also been identified for ultra-cold 
dusty plasmas \cite{Stenflo}--\cite{Misra}, where quantum effects can be used for plasma diagnostics. The most recent developments take care of spin 
effects in non relativistic quantum plasmas \cite{mm} as well as the associated magnetohydrodynamics equations \cite{mmm}, with possible important 
applications for solid state plasmas as well as in the vicinity of pulsars and magnetars. In addition, there are the analysis of the nonlinear 
instability of polaritons \cite{Shuklanew}, of the dynamics of dark solitons and vortices in quantum electron plasma \cite{Shukla}, of nonlinear interactions between intense circularly polarized electromagnetic waves and electron plasma oscillations \cite{SB}, of the behavior of quantum diodes \cite{Ang}, 
of nonlinear quantum dust-acoustic waves \cite{Wadati}, of quantum ion-acoustic double layers \cite{Moslem}, 
of linear and nonlinear ion-acoustic waves in unmagnetized electron-positron-ion quantum plasmas \cite{Al}, the construction of classes of solutions for 
the quantum Zakharov-Kuznetsov equation \cite{Mosl}, of electron-acoustic solitary waves in dense quantum electron-ion plasmas \cite{Misr} and of linear 
and nonlinear dust ion acoustic waves in ultra cold quantum dusty plasmas \cite{Khan}. Finally, there are new experimental studies \cite{Glenzer} of weakly degenerate quantum plasmas in a gaseous regime (i.e. non solid state plasmas).
A recent review on quantum plasma models and their range of validity can be found in \cite{Manfred}.

In some systems \cite{Suh}, the ultimate influence of quantum mechanics in plasmas is the stabilization of some classically unstable mode, 
for sufficiently strong quantum effects. However, in the intermediate regime where quantum effects are not too intense but are nevertheless not 
negligible, there are situations where unexpected quantum instabilities can arise. Examples on this are the quantum two-stream and three-stream
instabilities \cite{Haas}--\cite{bjp}, showing unstable modes of pure quantum nature and no classical counterpart. In addition, 
unlike classical plasmas, there is no Penrose functional determining the linear stability properties of 
quantum plasmas \cite{Haas2}. These considerations points to the subtle r\^ole
played in plasmas by quantum diffraction effects like tunneling and wave-packet spreading. Therefore, it is a relevant subject, to work out well known
classical instabilities now in the context of quantum plasma models. In this perspective, the present work considers Weibel's instability \cite{Weibel}.
Weibel instability arises from temperature anisotropy in the equilibrium distribution function and is one of the fundamental instabilities of
plasma physics. In the more recent years, it has been the central concept in several instances, like in fast ignitor scenarios \cite{Silva}, for 
particle acceleration and magnetic field generation in astrophysical settings \cite{Schli, Sch, Fonseca}, for 
collective non-Abelian Weibel instabilities in melting color glass condensates \cite{Romatschke}, in covariant relativistic scenarios \cite{Rolffs, Yoo}, in electron-positron relativistic shocks \cite{Nishikawa}, with kappa and generalized $(r,q)$ distributions \cite{Zaheer} and in laser heated plasmas \cite{Sangam}.

The aim of this contribution is to get detailed information about the influence of quantum effects on Weibel's instability. For this purpose, it is
used the (kinetic) Wigner-Maxwell model, which is the quantum counterpart of the Vlasov-Maxwell system. In this context, the Wigner function plays the
same r\^ole as the classical distribution function. Due to its mathematical complexity, frequently the electromagnetic Wigner equation \cite{HH} 
has not been taken as the basic tool in electromagnetic quantum plasmas. Rather, most works relies on the quantum hydrodynamic model \cite{HH, Manfredi2}. 
Here, however, the
kinetic description is used in order to provide a easier comparison with previous results on Weibel's instability, which were constructed in terms of
the Vlasov-Maxwell system. Nevertheless, notice that there are instances \cite{Basu} where fluid models were also applied to Weibel's instability.
Since the treatment is restricted to small amplitude waves, the electromagnetic Wigner equation can still provide meaningful results
without too complicated analytical difficulties. 

Sometimes, the Weibel instability is treated in conjunction with counter-streaming beams 
and/or ambient magnetic fields. However, as pointed out in \cite{Bret}, in these cases it is more 
appropriated to talk about filamentation instability. Indeed, the Weibel instability is prompted only by a single anisotropic system. So, here it is followed the original approach by Weibel \cite{Weibel}, focusing only on the consequences of temperature anisotropy, but now
allowing also for quantum effects. In addition, the analysis is restricted to non-relativistic systems. Notice that recently quantum effects were addressed for the filamentation (but not Weibel) instability \cite{Bre}. In this case, quantum effects have been shown to reduce both the unstable wave-vector domain and the maximum growth rate. 

This paper is organized as follows. In Section II, we write the dispersion relation for small amplitude transverse electromagnetic waves in quantum plasmas, as derived from the Wigner-Maxwell system. Then it is assumed a double Gaussian equilibrium distribution function where temperature anisotropy is allowed. The resulting dispersion relation is analyzed for several limiting cases in Section III. Namely, there are considered the ultra-quantum case, the semiclassic case for small wave-lengths, and the semiclassic case for long wave-lengths In Section IV an equilibrium with extreme temperature anisotropy and a three-dimensional water bag model are considered, allowing for more detailed analytic results. In all cases, quantum effects are stabilizing. Section V discuss possible applications in the case of neutron stars and white dwarfs as well as for tunnel-ionized plasmas. As is shown in the following, the quantum effects are enhanced for larger density and larger temperature anisotropy. For neutron stars and white dwarfs, the high densities tend to enhance quantum effects. 
For tunnel-ionized plasmas, the densities are not so high, but the temperature anisotropy can be significantly enough. 
Section VI is reserved to the conclusions. 

\section{Basic equations}
Consider a plasma composed of electrons (charge $-e$, mass $m$) and a neutralizing immobile ionic background. In terms of the
Wigner distribution function $f = f({\bf r},{\bf v},t)$, the electron particle density $n = n({\bf r},t)$ and the current density
${\bf J} = {\bf J}({\bf r},t)$ are given by
\begin{equation}
n = \int d{\bf v} f \,, \quad 
{\bf J} = - e\int d{\bf v}\,f\,{\bf v}  \,.
\end{equation}
All integrals are from minus to plus infinity unless otherwise stated. To proceed, it is 
necessary to work in terms of the electromagnetic potentials $(\phi({\bf r},t),{\bf A}({\bf r},t))$, 
since the electromagnetic Wigner equation \cite{HH, Arnold} is written in terms of them and
not the fields. In tangent space with coordinates $({\bf r},{\bf v})$, and time variable $t$, it reads
\begin{eqnarray}
\label{wign} 0 &=& 
\frac{\partial f}{\partial t} + (v_i - \frac{e A_i}{m})\left(\frac{\partial f}{\partial r_i} 
+ \frac{e}{m}\frac{\partial A_j}{\partial r_i}\frac{\partial f}{\partial v_j}\right) 
+ \frac{e}{m}\frac{\partial{\bf A}}{\partial t}\cdot\frac{\partial f}{\partial{\bf v}} + \\
&+& \frac{ie}{\hbar}\!\left(\frac{m}{2\pi\hbar}\right)^{3}\!\int\!\int\!d{\bf s}\,d{\bf v}'
\!e^{im\!({\bf v}\!-\!{\bf v}'\!)\cdot{\bf s}/\hbar}\!\left[\phi\!\left(\!{\bf r}\!+\!\frac{{\bf s}}{2}\right)\!-
\!\phi\!\left(\!{\bf r}\!-\!\frac{{\bf s}}{2}\right)\!\right]\!f({\bf r}\!,\!
{\bf v}') \nonumber \\
&-& \frac{ie^2}{2\hbar m}\!\left(\frac{m}{2\pi\hbar}\!\right)^{3}\!\int\!\int\!d{\bf s}\,d{\bf v}'\!e^{im\!({\bf v}\!-\!{\bf v}'\!)
\cdot{\bf s}/\hbar}\!\left[A^{2}\!\left({\bf r}\!+\frac{{\bf s}}{2}\right)\!-\!A^{2}\!\left({\bf r}\!-\frac{{\bf s}}{2}\right)
\!\right]\!f({\bf r}\!,\!{\bf v}')\nonumber\\
&+&\frac{e}{2m}\left(\frac{m}{2\pi\hbar}\right)^{3}\left[\frac{\partial}{\partial r_i} + \frac{e}{m}\frac{\partial A_j}{\partial r_i}
\frac{\partial}{\partial v_j}\right] \times \nonumber\\ &\times&\int\int d{\bf s}\,d{\bf v} e^{im({\bf v} - {\bf v}')\cdot{\bf 
s}/\hbar}\left[A_{i}\left({\bf r} + \frac{{\bf s}}{2}\right) + A_{i}\left({\bf r} - \frac{{\bf s}}{2}\right)\right]
f({\bf r},{\bf v}') \nonumber \\ 
&-&\!\!\!\!\frac{ie}{\hbar}\!\left(\frac{m}{2\pi\hbar}\!\right)^{3}\!\left(\!{\bf v}\!-\!\frac{e{\bf A}}{m}\!\right)
\!\cdot\!\int\!\!\int\!d{\bf s}\,d{\bf v}'\!e^{im({\bf v}\!-\!{\bf v}')\cdot{\bf s}/\hbar}\!\left[{\bf A}\!\left({\bf r}\!+\!\frac{{\bf s}}{2}\right)
\!-\!{\bf A}\!\left({\bf r}\!-\!\frac{{\bf s}}{2}\right)\!\right]\!f({\bf r}\!,\!{\bf v}') ,\nonumber
\end{eqnarray}
omitting the time-dependence of the several quantities and using summation convention in some terms. For the derivation of the electromagnetic Wig\-ner equation in this form, it was assumed the Coulomb gauge, $\nabla\cdot{\bf A} = 0$. Also, notice that 
the electromagnetic Wigner equation as in Eq. (3) of Ref. \cite{HH} was been written in phase space with canonical coordinates $({\bf q},{\bf p})$ and a
time variable $\tau \equiv t$. To derive (\ref{wign}), it is necessary to transform using ${\bf q} = {\bf r}$, ${\bf p} = m{\bf v} - e{\bf A}$ and
$\tau = t$, applying the chain rule, which implies, in particular,
\begin{eqnarray}
\frac{\partial}{\partial\tau} &=& \frac{\partial}{\partial t} + \frac{e}{m}\frac{\partial{\bf A}}{\partial t}\cdot\frac{\partial}{\partial{\bf v}} \,,\\
\frac{\partial}{\partial q_i} &=& \frac{\partial}{\partial r_i} + \frac{e}{m}\frac{\partial A_j}{\partial r_i}\frac{\partial}{\partial v_j} \,.
\end{eqnarray}
In the formal classical limit, $\hbar \rightarrow 0$, (\ref{wign}) reduces to the Vlasov equation, 
\begin{equation}
\frac{\partial f}{\partial t} + {\bf v}\cdot\nabla f + \frac{e}{m}\left(\frac{\partial\phi}{\partial r_i} + \frac{\partial A_i}{\partial t} + v_{j}[\frac{\partial A_i}{\partial r_j} - \frac{\partial A_j}{\partial r_i}]\right)\frac{\partial f}{\partial v_i} = 0 \,.
\end{equation}
As a last remark, here we also have corrected a mistyping in one of the signals at the third term in the right-hand side of Eq. (3) of Ref. \cite{HH}. This mistyping produce no harm for the conclusions of this reference. 

Now take linear wave propagation along the $O\hat{z}$ axis, with wave-vector ${\bf k} = k\hat{z}$, identically zero scalar potential and 
\begin{equation}
{\bf A} = {\bf A}_{\bot}\exp(i[kz-\omega t]) \,,
\end{equation}
where ${\bf A}_{\bot}$ is a first-order object satisfying $ {\bf k}\cdot{\bf A}_{\bot} \equiv 0$. This form for the vector potential is 
consistent with a transverse magnetic field. Also, the Wigner function expands as $f = f_0 + f_{1}\exp(i[kz-\omega t])$, where 
$f_0 = f_{0}({\bf v})$ is the equilibrium Wigner function, satisfying 
\begin{equation}
\label{e0}
\int d{\bf v} f_0 = n_{0} \,, \quad \int d{\bf v}\, f_0 \, {\bf v}  = 0 \,,
\end{equation}
where $n_0$ is the ambient ion density and $f_1$ is a first order perturbation. Accordingly, the
current density is given by
\begin{equation}
{\bf J} = - e \int d{\bf v} f_1 {\bf v} \,, 
\end{equation}
omitting the $\exp(i[kz-\omega t])$ dependence. 

In this context, Amp\`ere's law yields
\begin{equation}
\label{e1}
(\omega^2 - c^2 k^2) {\bf A}_{\bot} = \frac{e}{\varepsilon_0} \int d{\bf v} f_1 {\bf v} \,.
\end{equation}
The homogeneous Maxwell equations are identically satisfied
when working with the electromagnetic potentials, while Poisson equation is satisfied since it can be shown that 
there are no charge density fluctuations 
($\int\,d{\bf v} f_1 = 0$), as in the classical Weibel instability. Therefore, the only
remaining equation is the linearized Wigner equation. Using (\ref{wign}), the result is 
\begin{eqnarray}
(\omega &-& kv_{z})\left(f_1 - \frac{e}{m}{\bf A}_{\bot}\cdot\frac{\partial f_0}{\partial{\bf v}}\right)  \nonumber \\ \label{e2} &=&
\frac{e\,{\bf v}\cdot{\bf A}_{\bot}}{\hbar} \,\left(f_{0}(v_{x},v_{y},v_{z} - \frac{\hbar k}{2m}) - f_{0}(v_{x},v_{y},v_{z} +
\frac{\hbar k}{2m})\right)\,,
\end{eqnarray}
for ${\bf v} = (v_{x},v_{y},v_{z})$ and $\hbar = h/(2\pi)$ being the scaled Planck's constant. Combining (\ref{e0}), (\ref{e1}) and (\ref{e2}), assuming that the equilibria $f_0$ are even functions of all velocity components also satisfying the condition 
$f_{0}(v_{x},v_{y},v_z) = f_{0}(v_{y},v_{x},v_z)$ and doing an integration by parts,  
there follows the quantum dispersion relation for transverse waves
(${\bf k}\cdot{\bf E} = 0$),
\begin{eqnarray}
\omega^2 - \omega_{p}^2 &-& c^2 k^2 + \frac{m \omega_{p}^2}{2n_0 \hbar}\int d{\bf v}\left(\frac{v_{x}^2 + v_{y}^2}{\omega -
kv_z}\right) \times \nonumber \\ \label{e3} &\times& \left(f_{0}(v_{x},v_{y},v_{z} + \frac{\hbar k}{2m}) - f_{0}(v_{x},v_{y},v_{z} - \frac{\hbar k}{2m})\right)= 0 \,,
\end{eqnarray}
where $\omega_p = (n_{0}e^2/(m\varepsilon_{0}))^{1/2}$ is the plasma frequency. 

Consider the specific case of the equilibrium Wigner function 
\begin{equation}
f_0 = \frac{n_0}{T_{\parallel}^{1/2}T_{\bot}}\left(\frac{m}{2\pi}\right)^{3/2} \exp\left(-\frac{m}{2T_{\bot}}(v_{x}^2 + v_{y}^2) - \frac{m v_{z}^2}{2T_{\parallel}}\right) \,,
\end{equation}
showing temperature anisotropy between parallel and perpendicular directions. The temperatures $T_\bot$ and $T_{\parallel}$ are measured in terms of
energy units (Boltzmann's constant is $\kappa_B \equiv 1$). Then the dispersion relation (\ref{e3}) develops into
\begin{equation}
\label{e4}
\omega^2 - c^2 k^2 - \omega_{p}^2 \left(1 + \frac{T_\bot}{T_{\parallel}}W_Q\right) = 0 \,, 
\end{equation}
where 
\begin{equation}
W_Q = \frac{mv_{\parallel}}{2\hbar k}\left(Z(\frac{\omega}{kv_\parallel} + \frac{\hbar k}{2mv_\parallel}) - Z(\frac{\omega}{kv_\parallel} -
\frac{\hbar k}{2mv_\parallel})\right) \,,
\end{equation}
for $v_{\parallel} = (2T_{\parallel}/m)^{1/2}$ and where $Z$ is the plasma dispersion function. The function $W_Q$ can be appropriately named
a ``quantum Weibel function". Apart from a scale factor, it is a centered finite difference version of the derivative of the plasma dispersion
function. In the formal classical limit when $\hbar \rightarrow 0$, 
\begin{equation}
W_Q \rightarrow W\left(\frac{\omega}{kv_\parallel}\right) \equiv - 1 - \frac{\omega}{kv_\parallel}Z\left(\frac{\omega}{kv_\parallel}\right) \,,
\end{equation}
where the function $W(\omega/(kv_\parallel))$ is the same (Weibel) function as the one in Eq. (5) of Ref. \cite{Basu}. Hence, in this formal
classical limit, the dispersion relation (\ref{e4}) reduces to the classical one, in accordance with the correspondence principle. Notice that
quantum effects appears only through the non-dimensional parameter
\begin{equation}
\label{y}
H = \frac{\hbar k}{mv_\parallel} \,,
\end{equation}
basically depending only on the longitudinal quantities $k$ and $v_{\parallel}$. 

\section{Limiting cases}

We consider separately the ultra-quantum case, the small quantum parameter and wave-number case and the small quantum parameter and large wave-number
case.
\subsection{Ultra-quantum case ($H \gg 1$)}

It is interesting to check the behavior of the dispersion relation in the case of very large or very small quantum effects. In this regard, an 
useful alternative (exact) expression for $W_Q$ is
\begin{equation}
\label{e5}
W_Q = \frac{1}{2\sqrt{\pi}}\int\,\frac{d\xi e^{-\xi^2}}{(\xi - \frac{\omega}{kv_{\parallel}})^2 - H^2/4} \,,
\end{equation}
for $\xi = \omega/(kv_\parallel)$. Assuming large quantum effects, so that $H^2 \gg |\omega/(kv_{\parallel})|^2$, it
follows from (\ref{e5}) that $W_Q \simeq -2/H^2$, so that (\ref{e4}) imply
\begin{equation}
\omega^2 = c^2 k^2 + \omega^2_{p} \left(1 - \frac{2m^2 v_{\parallel}^2 T_\bot}{\hbar^2 k^2 T_{\parallel}}\right) \,.
\end{equation}
There can be instability ($\omega^2 < 0$), provided there is also sufficient temperature anisotropy,
\begin{equation}
\label{e6}
\frac{T_\bot}{T_\parallel} > \frac{H^2}{2}\left(1 + \frac{c^2 k^2}{\omega_{p}^2}\right) \,.
\end{equation}
However, since the right hand side of the last inequality is an increasing function of $H$, one conclude that quantum effects play an stabilizing r\^ole. 
Another way to derive this conclusion comes from rewriting (\ref{e6}) in terms of the instability condition
\begin{equation}
\label{x}
k^2 < k_{c}^2 \equiv \frac{\omega_{p}^2}{2c^2}\left(-1 + (1 + \frac{16m c^2 T_\bot}{\hbar^2 \omega_{p}^2})^{1/2}\right) \,.
\end{equation}
For in\-crea\-sing quan\-tum ef\-fects, the critical wa\-ve-\-num\-ber $k_c$ be\-co\-mes smal\-ler, eventually dropping to zero. 
Notice also that (\ref{x}) and $H^2 \gg 1$ are conflicting conditions. 
A detailed but cumbersome analysis shows that both conditions can be satisfied only for $\hbar\omega_{p}/(mc^2) \gg 1$, 
violating the non-relativistic assumption of the present model.  

\subsection{Semiclassic case at small wave-lengths ($H \ll 1$ and $|\xi| \ll 1$)} 
Now retaining only the first-order quantum correction, one get
\begin{equation}
\label{e7}
W_Q = - 1 - \xi Z(\xi) + \frac{H^2}{12}\left(2 + 3\xi Z(\xi) - 2\xi^2 - 2\xi^3 Z(\xi)\right) + O(H^4) \,.
\end{equation}
For such small quantum effects and also taking small wave-lengths so that $|\xi| \ll 1$ and $Z(\xi) \simeq i\sqrt{\pi}$, there follows from
(\ref{e7}) that
\begin{equation}
W_Q \simeq - 1 - i\xi\sqrt{\pi} + \frac{H^2}{6} \,,
\end{equation}
while the dispersion relation (\ref{e4}) produces
\begin{equation}
\label{e8}
\omega = \frac{ikv_\parallel T_\parallel}{\sqrt{\pi}T_\bot}\left(\frac{T_\bot}{T_\parallel}(1 - \frac{H^2}{6}) - 1 - \frac{c^2 k^2}{\omega_{p}^2}\right) \,.
\end{equation}
In the derivation of (\ref{e8}), it was used $|\omega^2|/(c^2 k^2) = |\xi^2| v_{\parallel}^2/c^2 \ll 1$, also consistent with the non-relativistic
approximation. Assuming positive wave-numbers, it is apparent from (\ref{e8}) that purely growing waves (frequencies composed only by a positive
imaginary part) can exist for sufficiently large temperature anisotropy, and that this necessary anisotropy
becomes larger for increasing $H$. Specifically, for small quantum effects and wave-lengths, there will be a purely growing wave only if
\begin{equation}
\frac{T_\bot}{T_\parallel} > \frac{1}{1-H^2/6} \simeq 1 + \frac{H^2}{6} \,,
\end{equation}
showing the need of extra anisotropy for instability, due to quantum effects. In addition, in this combined regime of small quantum effects and
wave-lengths, it can be proven that the unstable wave-numbers are restricted to 
\begin{equation}
\label{kc}
k^2 < k_{c}^2 \equiv \frac{\omega_{p}^{2}(T_{\bot}/T_\parallel - 1)}{c^2 \left(1 + \frac{T_\bot \hbar^2 \omega_{p}^2}{12 T_{\parallel}^2 m
 c^2}\right)} \,.
\end{equation}
Once again, the stabilizing nature of (now small) quantum effects is apparent, since the allowable unstable wave-numbers occurs for a smaller
range for increasing $H^2$. 

\subsection{Semiclassic case at large wave-lengths ($H \ll 1$ and $|\xi| \gg 1$)} 
From the expansion (\ref{e7}) and using $Z(\xi) \simeq - 1/\xi - 1/(2\xi^3)$ when $|\xi| \gg 1$, for large wave-lengths, it follows the dispersion
relation
\begin{equation}
\label{e9}
\omega^2 - c^2 k^2 - \omega_{p}^2 \left(1 + \frac{k^2 T_{\bot}}{m\omega^2}(1 - \frac{H^2}{4})\right) = 0 \,.
\end{equation}
For $|\omega| \ll ck$, (\ref{e9}) yields the purely growing mode
\begin{equation}
\label{e10}
\omega = ik\left(\frac{T_{\bot}(1 - H^2/4)}{m}\right)^{1/2} \left(\frac{\omega_{p}^2}{c^2 k^2 + \omega_{p}^2}\right)^{1/2} \,.
\end{equation}
Notice that the growth rate becomes smaller for larger $H$. Also, in view of (\ref{e10}), the condition $|\xi| \gg 1$ can be attained only for $T_\bot 
\gg T_{\parallel}/(1 - H^2/4)$. Equations (\ref{e9}) and (\ref{e10}) are formally the same as Eqs. (8) and (9) of Ref. \cite{Basu}, making the replacement $T_\bot \rightarrow T_{\bot}(1 - H^2/4)$, indicating the need of extra temperature anisotropy, in view of quantum effects. 

\section{Toy models and more detailed analytical results}
Anisotropic Gaussian distributions are amenable to analytical results only for limiting situations. This Section consider some toy models which 
behave more friendly in this respect. As a first example where full analytical results are available, consider the equilibrium Wigner function
\begin{equation}
f_0 = \frac{n_0 m}{2\pi T_\bot} \delta(v_z) \exp\left(-\frac{m}{2T_{\bot}}(v_{x}^2 + v_{y}^2)\right) \,,
\end{equation}
which can be viewed as a distribution with extreme temperature anisotropy ($T_\parallel \rightarrow 0$). Inserting into (\ref{e3}) and proceeding as before, it results
\begin{equation}
\omega^2 = \frac{1}{2}\left(\omega_{p}^2 + c^2 k^2 + \frac{\hbar^2 k^4}{4m^2} \pm \left[(\omega_{p}^2 + c^2 k^2 - \frac{\hbar^2 k^4}{4m^2})^2 + \frac{4k^2 \omega_{p}^2 T_\bot}{m}\right]^{1/2}\right) \,.
\end{equation}
One of the roots is unstable ($\omega^2 < 0$), provided
\begin{equation}
k^2 < k_{c}^2 \equiv \frac{\omega_{p}^2}{2c^2}\left(1 + \frac{16 mc^2 T_\bot}{\hbar^2 \omega_{p}^2}\right)^{1/2} -  \frac{\omega_{p}^2}{2c^2} \,,
\end{equation}
showing, once again, stabilization due to increasing quantum effects. 

The classical transverse Weibel instability has sometimes considered in terms of water bag distributions \cite{Silva, Bret, Yoon}. As a second
example, also amenable to detailed calculations, take the following three-dimensional water bag \cite{Bret} equilibrium,
\begin{eqnarray}
f_0 = \frac{n_0}{8 v_{\bot}^2 v_\parallel} \left(\theta(v_x + v_\bot) - \theta(v_x - v_\bot)\right) \left(\theta(v_y + v_\bot) - \theta(v_y - v_\bot)\right) \times 
\nonumber \\   \times \left(\theta(v_z + v_\parallel) - \theta(v_z - v_\parallel)\right) \,, 
\;\;\;\;  \;\;\;\;  \;\;\;\;  \;\;\;\;  \;\;\;\;   
 \strut 
\end{eqnarray}
where $\theta$ is the Heaviside function and in this context 
$v_\bot$ and $v_\parallel$ are related to dispersion of velocities in the perpendicular plane and along
the $O\hat{z}$ axis, as before. Then the dispersion relation (\ref{e3}) yields
\begin{equation}
\label{e11}
\omega^2 - \omega_{p}^2 - c^2 k^2 = \frac{m\omega_{p}^2 v_{\bot}^2}{6\hbar kv_\parallel} \ln\left(\frac{\omega^2 - (kv_\parallel -
\hbar k^2/(2m))^2}{\omega^2 - (kv_\parallel + \hbar k^2/(2m))^2}\right) \,,
\end{equation}
not in polynomial form as in the formal classical limit \cite{Bret} for the corresponding equilibrium. However, some analytical results are
still available. Assuming purely growing instabilities with a growth rate $\gamma$, so that $\omega = i\gamma$, and also disregarding $\gamma^2$ at
the left-hand side of (\ref{e11}) assuming $\gamma^2 \ll \omega_{p}^2$, there follows
\begin{equation}
\label{e12}
\gamma^2 = \frac{\hbar k^3 v_\parallel}{m} \coth\left(\frac{3\hbar kv_\parallel (c^2 k^2 + \omega_{p}^2)}{m\omega_{p}^2 v_{\bot}^2}\right) -
k^2 v_{\parallel}^2 - \frac{\hbar^2 k^4}{4m^2} \,.
\end{equation}
Notice that (\ref{e12}) is not valid in the formal classical limit $\hbar \equiv 0$ because then the associated growth rate would not be small.
Rather, in this limit one has to Taylor expand the right-hand side of (\ref{e11}), and then the results from \cite{Bret} are recovered. 

To proceed, it is convenient to adopt the following rescaling,
\begin{equation}
\bar{\gamma} = \frac{\gamma}{\omega_p} \,, \quad \bar{k} = \frac{kv_\parallel}{\omega_p} \,, \quad \bar{H} = \frac{\hbar\omega_p}{mv_{\parallel}^2} \,,
\quad \bar{v}_\bot = \frac{v_\bot}{c} \,, \quad \bar{v}_\parallel = \frac{v_\parallel}{c} \,, 
\end{equation}
so that (\ref{e12}) becomes
\begin{equation}
\label{e13}
\bar{\gamma}^2 = \bar{H}\bar{k}^3 \coth\left(
\frac{3\bar{H}\bar{k}}{\bar{v}_{\bot}^2}(\bar{k}^2 + \bar{v}_{\parallel}^2)
\right) - \bar{k}^2 - 
\frac{\bar{H}^2 \bar{k}^4}{4} \,.
\end{equation}
It is possible to estimate the maximum wave-number for instability, using $\coth(\xi) \simeq 1/\xi$ for $|\xi| \ll 1$. Assuming that this expansion 
is valid, one will conclude from (\ref{e13}) that $\bar{\gamma}^2 > 0$ provided
\begin{equation}
\label{e14}
\bar{k}^2 < \bar{k}_{m}^2 \equiv \frac{1}{2\bar{H}^2}\left([(\bar{H}^2 \bar{v}_{\parallel}^2 - 4)^2 + \frac{16\bar{H}^2 \bar{v}_{\bot}^2}{3}]^{1/2} -
\bar{H}^2 \bar{v}_{\parallel}^2 - 4\right) \,.
\end{equation}
From (\ref{e14}), it can be shown that there will exist some unstable mode ($\bar{k}_{m}^2 > 0$) if and only if there
is sufficient temperature anisotropy, $\bar{v}_\bot > \sqrt{3}
\bar{v}_\parallel$, independently of the strength of the quantum effects, and in accordance with the scaling found for classical plasma \cite{Bret}.
However, it can be deduced from (\ref{e14}) that the critical wave-number shrinks to zero as $\bar{H} \rightarrow \infty$. 

%Figure \ref{figure1}  shows the behavior of $\bar{\gamma}^{2}(\bar{k})$ in terms of several values of $\bar{H}$ according to (\ref{e13}), for fixed values
%$\bar{v}_\bot = 0.7$ and $\bar{v}_\parallel = 0.1$. One sees that for increasing quantum effects there are smaller maximal growth rates and smaller
%maximal wave-numbers for instability. Once again, quantum effects plays a stabilizing r\^ole. In addition, Figure \ref{figure2} shows the behavior of
%$\bar{\gamma}^{2}(\bar{k})$ in terms of several values of $\bar{v}_\bot$ according to (\ref{e13}), for fixed values $\bar{H} = 1.0$ and
%$\bar{v}_\parallel = 0.1$. One sees that for decreasing anisotropy (smaller values of $\bar{v}_\bot$) the instability occurs for smaller regions
%in $k$ space and with smaller maximal growth rates. 

%\begin{figure*}[ht]
%\begin{center}
%\leavevmode
%{\par\centering
%\resizebox*{0.80\textwidth}{!}{\rotatebox{0}{\includegraphics{gamma2_H.eps}}}
%\par}
%\caption{$\bar{\gamma}^{2}(\bar{k})$ for $\bar{H}=1$ (solid line), $\bar{H} = 10$ (dashed line) and $\bar{H} = 25$ (dotted line), for $\bar{v}_\parallel = 0%.1$ and for $\bar{v}_\bot = 0.7$.}
%\label{figure1}
%\end{center}
%\end{figure*}

%\begin{figure*}[ht]
%\begin{center}
%\leavevmode
%{par\centering
%resizebox*{0.80\textwidth}{!}{\rotatebox{0}{\includegraphics{gamma2_vz.eps}}}
%par}
%\caption{$\bar{\gamma}^{2}(\bar{k})$ for $\bar{v}_\bot = 0.3$ (solid line), $\bar{v}_\bot = 0.5$ (dashed line) and $\bar{v}_\bot = 0.7$ (dotted line), 
%for $\bar{H} = 1$ and for $\bar{v}_\parallel = 0.1$.} 
%\label{figure2}
%\end{center}
%\end{figure*}

\section{Applications}
Consider now the case of anisotropic Maxwellian equilibria. For semiclassic and small wave-lengths conditions (Section III.2), one can use (\ref{e8}) 
to show that the maximal value of the growth rate happens at $k = k_m = k_{c}/\sqrt{3}$, with $k_c$ given by (\ref{kc}). The 
corresponding maximal growth rate is then 
\begin{equation}
\label{gm}
\gamma_m = \left(\frac{8}{27\pi}\right)^{1/2} \omega_p \left(\frac{T_\parallel}{mc^2}\right)^{1/2} \frac{T_\parallel}{T_\bot} 
\frac{(T_{\bot}/T_\parallel - 1)^{3/2}}{\left(1 + \frac{T_{\bot}\hbar^2 \omega_{p}^2}{12 T_{\parallel}^2 mc^2}\right)^{1/2}} \,.
\end{equation}
The calculations leading to (\ref{gm}) were semiclassic in the sense that the parameter $H = \hbar k/mv_\parallel$ is taken as small. However, 
there can be significant deviations from the classical expression, coming from the $\hbar^2$ term in the denominator of (\ref{gm}), provided there 
is sufficient temperature anisotropy. These deviations are also enhanced by large densities. 
For instance, consider white dwarfs and neutron stars,  with 
typical values $n_0 \sim 10^{32} m^{-3}$, $T_\bot = 10^7 K$, $T_\parallel = T_\bot/100$. The origin of the 
temperature anisotropy can be, for instance, the propagation of a shock wave. For these parameters, for hydrogen plasma, 
one finds that the quantum corrected maximal growth rate is about $11 
\%$ smaller than the classical one. However, these calculations have to be taken with care, because they suppose that $H^2$ from (\ref{y}) at $k = k_m$ 
and $|\xi| = \gamma_{m}/(k_{m}v_{\parallel})$ are small quantities. For the chosen parameters one get $H^2 \sim 0.41$ and $|\xi| \sim 0.37$. For larger 
densities, $H^2$ would be even greater. 

Another interesting system where quantum corrections for Weibel instability 
can be significant are tunnel-ionized plasmas with negligible longitudinal temperature and 
where the perpendicular temperature can be controlled by a varying laser polarization. It has been argued \cite{Leemans} that the Weibel instability could 
be a mechanism for further increase of $T_\parallel$ with time. For a typical value \cite{Leemans} of $T_\parallel \sim 1 eV$,
one find
\begin{equation}
\label{v}
\frac{T_{\bot}\hbar^2 \omega_{p}^2}{12 T_{\parallel}^2 mc^2} =  2.3 \times 10^{-34} n_0 T_\bot \,,
\end{equation}
using SI units for $n_0$ and $T_\bot$. Although it is not easy to get large values for the quantity at the right-hand side of (\ref{v}), the 
fast progress in next generation intense laser-solid density plasma interaction experiments can be such that quantum effects stops 
the Weibel instability, specially for ultra-dense systems. For the largest densities \cite{Glenzer, Malkin} feasible now,
$n_0 \sim 10^{29} m^{-3}$, and for $T_\bot \sim 100 eV$, the right-hand side of (\ref{v}) has already a significant 
value of the order of $25$. Another way to get even larger quantum effects is the use of smaller values of $T_\parallel$. 
The results of this section, however, have to be taken with care, since they deal with dense plasmas which should be more properly treated
by Fermi-Dirac statistics. 

\section{Conclusions}
In general terms, quantum effects produces smaller growth rates and smaller ranges for unstable wave-numbers, in
the case of equilibria with distribution functions anisotropic in temperature. Some applications in astrophysical scenarios and in tunnel-ionized plasma 
were discussed. Eventually, in the ultra-quantum case, the unstable region shrinks to
zero.  We can understand this result in an heuristic way as follows. Due to wave-particle spreading and tunneling, quantum effects tends to enhance the
dispersion of particles in phase space. This corresponds to an effectively smaller temperature anisotropy, or thermalization, so that the original
ratio $T_{\bot}/T_{\parallel}$ have to be greater to produce the same instability results as in classical plasma. Similar spreading in phase space also 
occurs in the case of quantum corrected Bernstein-Greene-Kruskal modes \cite{Luque}. However, as mentioned in the
Introduction, sometimes quantum effects can also gives unexpected enhancement of plasma instabilities. Therefore, it is interesting to pursue this 
trend, looking at the behavior of additional well-known classical plasma instabilities, in the context of quantum plasma models. Also, an 
important issue is the inclusion of relativistic and spin effects. 

\vskip .5cm
{\bf Acknowledgements}
\vskip .5cm

The author acknowledges the support provided by a fellowship of the Alexander von Humboldt Foundation.

\end{document}